\documentclass[12pt]{article}
\usepackage{hyperref}
\hypersetup{
colorlinks=true,
linkcolor=black,
citecolor=black,
urlcolor=cyan,
filecolor=black,
pdfborder={0 0 0},
}

\usepackage{amssymb}
\usepackage{amsfonts}
\usepackage{mathrsfs}
\usepackage{latexsym}
\usepackage{amsmath}
\usepackage{graphics}
\usepackage{graphicx}
\usepackage{sectsty}
\usepackage{setspace}
\usepackage{multirow}
\sectionfont{\normalsize\bf}
\subsectionfont{\rm\normalsize}
\def\be{\begin{equation}}
\def\ee{\end{equation}}
\def\bal{\begin{align}}
\def\eal{\end{align}}
\def\bd{\left|\begin{matrix}}
\def\ed{\end{matrix}\right|}
\def\bm{\left[\begin{matrix}}
\def\ema{\end{matrix}\right]}
\input amssym.def
\input amssym
\def\Rop{{\Bbb R}}
\def\Cop{{\Bbb C}}
\def\Pop{{\Bbb P}}

\def\O{{\cal O}}
\def\A{{\cal A}}
\def\R{{\cal R}}
\def\V{{\cal V}}
\def\N{{\cal N}}
\def\P{{\cal P}}
\def\F{{\cal F}}
\def\H{{\cal H}}
\def\bpi{{\bar\pi}}
\def\half{{\scriptstyle {1\over 2}}}

\numberwithin{equation}{section}

\begin{document}

\thispagestyle{empty}
\begin{flushright}
\end{flushright}
\baselineskip=16pt
\vspace{.5in}
{
\begin{center}
{\bf The Polynomial Form of the Scattering Equations}
\end{center}}
\vskip 1.1cm
\begin{center}
{Louise Dolan}
\vskip5pt

\centerline{\em Department of Physics}
\centerline{\em University of North Carolina, Chapel Hill, NC 27599} 
\bigskip
\bigskip        
{Peter Goddard}
\vskip5pt

\centerline{\em School of Natural Sciences, Institute for Advanced Study}
\centerline{\em Princeton, NJ 08540, USA}
\bigskip
\bigskip
\bigskip
\bigskip
\end{center}

\abstract{\noindent 
The scattering equations, recently proposed by Cachazo, He and Yuan as providing a kinematic basis for describing tree amplitudes for massless particles in arbitrary space-time dimension (including scalars, gauge bosons and gravitons), are reformulated in polynomial form. The scattering equations for $N$ particles are shown to be  equivalent to a M\"obius invariant system of $N-3$  equations, $\tilde h_m=0$, $2\leq m\leq N-2$, in $N$ variables, where $\tilde h_m$ is a homogeneous polynomial of degree $m$, with the exceptional property of being linear in each variable taken separately. Fixing the M\"obius invariance appropriately, yields polynomial equations  $ h_m=0$, $1\leq m\leq N-3$,  in $N-3$ variables, where $h_m$ has degree $m$. The linearity of the equations in the individual variables  facilitates computation, {\it e.g.} the elimination of variables to obtain single variable equations determining the solutions. Expressions are given for the tree amplitudes in terms of the $\tilde h_m$ and $h_m$. The extension to the massive case for scalar particles is described and the special case of four dimensional space-time is discussed.
}
\bigskip

\setlength{\parindent}{0pt}
\setlength{\parskip}{6pt}

\setstretch{1.05}
\vfill\eject
\vskip50pt
\section{Introduction}
\label{Introduction}

In this paper, we study the scattering equations recently proposed by Cachazo, He and Yuan (CHY) \cite{CHY0} as a basis for describing the kinematics of massless particles,
\be
f_a(z,k)=0,\quad a\in A,\qquad \hbox{where}\quad  f_a(z,k)=\sum_{b\in A\atop b\ne a}{k_a\cdot k_b\over z_a-z_b},\label{SE}
\ee
and where $k_a$ are the momenta of the particles labeled by $a\in A$, with $N=|A|$, the number of elements in $A$, and $k_a^2=0$.  Typically, we shall take $A=\{1,2,\ldots,N\}$.

Our principal result is that these equations are equivalent to the homogeneous polynomial equations
\be
\sum_{S\subset A\atop |S|=m}k_S^2\;z_S=0, \qquad 2\leq m\leq N-2, \label{PSE}
\ee
where the sum is over all $N!/m!(N-m)!$ subsets $S\subset A$ with $m$ elements, and 
\be
k_S=\sum_{b\in S\atop }k_b,\qquad z_S=\prod_{a\in S}z_a, \qquad S\subset A. \label{kS}
\ee
 Note that the coefficients, $k_S^2, S\subset A$, in (\ref{PSE}) are precisely all the  
 Mandelstam variables, {\it i.e.}  the denominators of propagators of tree diagrams for processes involving the particles with momenta $k_a, a\in A$. 

We first recall basic properties of the equations (\ref{SE}) \cite{CHY0}--\cite{DG1}. The equations (\ref{SE}) are invariant under M\"obius transformations,
\be 
z_a\mapsto \zeta_a={\alpha z_a+\beta\over\gamma z_a+\delta},\quad a\in A,\label{Mt}
\ee
in that, if $z=(z_a)$ is a solution to (\ref{SE}), $\zeta=(\zeta_a)$ also provides a solution, $ f_a(\zeta,k)=0, a\in A$, provided that momentum is conserved and the particles are massless, for, using momentum conservation,
\be
\sum_{b\in A\atop b\ne a}{k_a\cdot k_b\over \zeta_a-\zeta_b}=\sum_{b\in A\atop b\ne a}{(cz_a+d)(cz_b+d)k_a\cdot k_b\over (ad-bc)(z_a-z_b)}
={(cz_a+d)^2\over (ad-bc)}f_a(z,k)+{c(cz_a+d)\over (ad-bc)}k_a^2,
\ee
which vanishes if $f_a(z,k)=0$ and $k_a^2=0$.

The M\"obius invariance implies that only $N-3$ of the equations (\ref{SE}) are independent. Relations between them are conveniently derived by noting that, if 
\be 
U(z,k)=\prod_{a<b}(z_a-z_b)^{-k_a\cdot k_b}, \quad\hbox{then}\quad  {\partial U\over\partial z_a}=-f_aU.
\ee
and the M\"obius invariance of the system of equations (\ref{SE}) can be deduced from that of $U(z,k)$ \cite{DG1}. Then, considering an infinitesimal M\"obius transformation, $\delta z_a=\epsilon_1+\epsilon_2z_a+\epsilon_3z_a^2$, $\delta U=0$
implies the identities:
\be 
\sum_{a\in A} f_a(z,k)=0;\qquad\sum_{a\in A}z_a f_a(z,k)=0;\qquad\sum_{a\in A}z_a^2 f_a(z,k)=0.\label{ids}
\ee

So only $N-3$ of the conditions (\ref{SE}) are needed to restrict to its set of solutions. This can be achieved by the delta function
\be
{\prod_{a\in A\atop }}'\delta\left( f_a(z,k)\right)\equiv (z_i-z_j)(z_j-z_k)(z_k-z_i)\prod_{a\in A\atop a\ne i,j,k}\delta\left( f_a(z,k)\right)\label{prodp}
\ee
which is independent of the choice of $i,j,k\in A$. Under the M\"obius transformation (\ref{Mt}), 
\be
{\prod_{a\in A\atop }}'\delta\left( f_a(z,k)\right)\mapsto {\prod_{a\in A\atop }}'\delta\left( f_a(\zeta,k)\right)=\prod_{a\in A\atop }{\alpha\delta-\beta\gamma\over (\gamma z_a+\delta)^2}{\prod_{a\in A\atop }}'\delta\left( f_a(z,k)\right),\label{prodp}
\ee
so that the integrand of 
\be
\A_N=\int \Psi_N(z,k){\prod_{a\in A\atop }}'\delta\left( f_a(z,k)\right)\prod_{a\in A}{dz_a\over (z_a-z_{a+1})^2}\bigg/d\omega,\label{famp}
\ee
where the invariant measure on the M\"obius group,
\be
d\omega={dz_rdz_sdz_t\over (z_r-z_s)(z_s-z_t)(z_t-z_r)}\;, \label{Mm}
\ee
is M\"obius invariant provided that the function $\Psi_N(z,k),$ which may also depend on polarizations, is itself M\"obius invariant. [Note that, in (\ref{famp}), a cyclic ordering on $A$ has been chosen, with $a+1$ following $a$, say $(1,2,\ldots,N)$ with $N+1\equiv 1$.] 
In order to give a precise interpretation of (\ref{famp}) that includes a sum over all, possibly complex, solutions of the  equations (\ref{SE}),  CHY rewrote the expression  as a contour integral 
\be
\A_N=\oint_\O\Psi_N(z,k){\prod_{a\in A\atop }}'{1\over f_a(z,k)}\prod_{a\in A}{dz_a\over (z_a-z_{a+1})^2}\bigg/d\omega\label{amp}
\ee
where the contour $\O$ encloses all the solutions of (\ref{SE}) and ${\prod}'$ is defined as in (\ref{prodp}).

CHY initially proposed a form of $\Psi_N(z,k)$ in (\ref{famp}) for tree amplitudes in pure Yang-Mills theory and in gravity in arbitrary space-time dimension \cite{CHY1}, subsequently noting that taking $\Psi_N$ constant would yield massless scalar $\phi^3$ theory \cite{CHY2}. This result was proved for massless $\phi^3$ and pure Yang-Mills theory in \cite{DG1}, where it was noted that the scattering equations (\ref{SE}) could be extended to the massive case, $k_a^2=\mu^2$, and that 
(\ref{amp}) then produced the tree amplitudes of massive $\phi^3$ for $\Psi_N$ constant.

It seems that the scattering equations (\ref{SE}) were first written down by Fairlie and Roberts \cite{FR}, who, before the string-theory interpretation of dual resonance models had been established in detail, were seeking a variation of the Veneziano model that was free of tachyons. They subsequently occurred in the work of Gross and Mende on the high energy behavior of string theory \cite{GM}. Fairlie and Roberts replaced the Fubini-Veneziano position field, $X^\mu(z),$ and the corresponding momentum field, $P^\mu(z),$
\be
X^\mu(z) = x^\mu -ip^\mu\log z + i\sum_{n\ne0}{a_n^\mu\over n}z^{-n},\qquad P^\mu(z)=i{dX^\mu(z)\over dz}=\sum_{n}a_n^\mu z^{-n-1},\quad a_0^\mu=p^\mu,
\ee
   expanded in terms of an infinite number of harmonic modes, $a_n^\mu$, with  
\be  
x^\mu(z) = -i\sum_{a\in A} k^\mu_a \log(z-z_a),\qquad p^\mu(z)=i{dx^\mu(z)\over dz}=\sum_{a\in A} {k^\mu_a \over z-z_a}.\label{xp}
\ee
The singularities of $x^\mu(z), p^\mu(z)$ on the Riemann sphere are evidently only at $z=z_a$ , provided that we have momentum conservation, $\sum_ak_a=0$. [If not, there will be a singularity at $z=\zeta^{-1}=\infty$, $x\sim -ik_\infty\log\zeta$, where $k_\infty=-\sum_a k_a$.] Fairlie and Roberts replaced the Virasoro conditions, which can be written
$\langle :P(z)^2:\rangle = \hbox{constant}$, with
\be
p(z)^2=\sum_{a,b}{k_a\cdot k_b \over (z-z_a)(z-z_b)}=0.\label{p2}
\ee
Given that $p(z)^2$ vanishes sufficiently fast at infinity, it vanishes everywhere  provided it has no singularities at $z=z_a,$ for $a\in A$. The absence of double poles requires $k_a^2=0, a\in A$, and the absence of simple poles implies (\ref{SE}). In an approach anticipating in general terms that of \cite{CHY1,CHY2}, they sought to define their model as a sum over the solutions of the scattering equations (\ref{SE}), rather than as an integral as in the Veneziano model. 

An important property of the scattering equations (\ref{SE}) is the way they factorize when one of the Mandelstam variables, $k^2_S,$ vanishes.  If $k_S^2\ne 0$ for all subsets $S$ with between $2$ and $N-2$ elements, the values of the $z_a, a\in A,$ are distinct. To demonstrate this, suppose that, as the $k_a$ vary (maintaining momentum conservation and the massless conditions),  two or more of the $z_a$ to tend to the same value, $z_0$. Suppose $z_a= z_0 +\epsilon x_a+\O(\epsilon^2)$ for $a\in S$, and $z_a\not\rightarrow z_0$ for $a\notin S$, as $\epsilon\rightarrow 0$; then the scattering equations factorize into two sets:
\be
 f_a(z,k)={1\over\epsilon}f_a^S(x,k)\left[1+\O(\epsilon)\right],\quad f_a^S(x,k)=\sum_{b\in S\atop b\ne a}{k_a\cdot k_b\over  x_a- x_b},\quad a\in S, \label{SES}
\ee
\be
 f_a(z,k)=\sum_{b\in S\atop }{k_a\cdot k_b\over z_a-z_0}+\sum_{b\notin S\atop b\ne a}{k_a\cdot k_b\over z_a-z_b}+\O(\epsilon),\quad a\notin S, \label{SES1}
\ee
implying in the limit that $f^S_a(x,k)=0, a\in S,$ and
\be
k_S^2=2\sum_{a,b\in S\atop b\ne a}{x_a\over  x_a- x_b}k_a\cdot k_b=2\sum_{a\in S}x_a\sum_{b\in S\atop b\ne a}{k_a\cdot k_b\over  x_a- x_b}=2\sum_{a\in S}x_af^S_a(x,k)=0.
\ee
In what follows, in general, we shall assume $k_S^2\ne 0$  for all subsets $S$ with between $2$ and $N-2$ elements and, consequently, that $z_a\ne z_b$ for $a\ne b$, $a,b\in A$.

In section \ref{Polynomial}, we derive the polynomial form (\ref{PSE}) of the scattering equations, first directly and then, more elegantly, by using their equivalence to (\ref{p2}). We then show show that the polynomial form possesses a factorization property corresponding to (\ref{SES}) and (\ref{SES1}). For the special kinematic configuration in which $k_S^2$ is independent of $S\subset A$ if $|S|=m$, the solutions of (\ref{PSE}) are given by complex roots of unity, which we describe in section \ref{Solutions}, where we also show how the fact that  (\ref{PSE}) is linear in each $z_a$ taken separately enables simple forms for single-variable equations to be found which determine the solutions to the scattering equations, at least for $N\leq 6$. Expressions for the scattering amplitudes in terms of the polynomials  (\ref{PSE}) are given in section \ref{Amplitudes}.
The M\"obius invariance of the integrand is demonstrated in section \ref{Moebius}, where it shown how the requirement of M\"obius invariance on equations of the form 
\be
\sum_{S\subset A\atop |S|=m}\lambda_S\;z_S=0, \qquad 2\leq m\leq N-2, \label{PSE2}
\ee
implies that $\lambda_S=k_S^2$, for suitable light-like momenta $k_a, a\in A$. The extension to the massive case is given in section \ref{Massive}, and special features of the four-dimensional space are discussed in section \ref{FourDim}. Some comments and indications of further directions for investigation are given in section \ref{Comments}.

\section{Polynomial Form for the Scattering Equations}
\label{Polynomial}

{\it The Polynomial Equations.} Studies of the scattering equations indicate that, after using M\"obius invariance to fix three of the $z_a$,  they have $(N-3)!$ solutions \cite{CHY0,CHY1}. B\'ezout's Theorem states that number of solutions to a system of polynomial equations, possessing a finite number of solutions, is bounded by the product of their degrees and, suitably counted, the number will attain this bound (see, {\it e.g.} \cite{Shaf}). This suggests that the scattering equations (\ref{SE})  should be equivalent to $N-3$ homogeneous equations in the $z_a$, $h_m=0, \;1\leq m\leq N-3,$ where $h_m$ has degree $m$. 

Writing
\be
g_{m}(z,k)=\sum_{a\in A}z_a^{m+1} f_a(z,k),\label{defg}
\ee
in (\ref{ids}), we noted that M\"obius invariance implies that $g_{-1}, g_0$ and $g_1$ vanish identically. For other integral values of $m$, (\ref{defg}) defines an homogeneous polynomial of degree $m$ in the $z_a$,
\begin{align}
g_m(z,k)&=\sum_{a,b\in A\atop a\ne b}{k_a\cdot k_bz_a^{m+1}\over z_a-z_b}
={1\over 2}\sum_{a,b\in A\atop a\ne b}k_a\cdot k_b{z_a^{m+1}-z_b^{m+1}\over z_a-z_b}\cr
&={1\over 2}\sum_{a,b\in A\atop a\ne b}k_a\cdot k_b\sum_{r=0}^{m}z_a^rz_b^{m-r} .\label{gdef}
\end{align}
The $N\times N$ matrix $Z_{ma}=z_a^{m+1}, a\in A, -1\leq m\leq N-2,$ is nonsingular, given that  $z_a\ne z_b$ for $a\ne b$, because $\det Z$ is the Vandermonde determinant,
\be
\det Z=\prod_{1\leq a<b\leq N}(z_a-z_b).
\ee
Thus the $N-3$ equations 
\be g_m=0,\qquad  \hbox{where}\quad  2\leq m\leq N-2, \label{gm}
\ee
 are equivalent to the equations $f_a=0, a\in A,$ $N-3$ of which are independent.
Since $g_m=0$, for all integers $m$, follows from  $f_a=0, a\in A,$ and these equations themselves follow from the set (\ref{gm}), it follows
that this set implies $g_n=0$ for all integers $n$.

The set  of $N-3$ homogeneous polynomial equations (\ref{gm}) is equivalent to the scattering equations (\ref{SE}) but this form is not convenient, {\it e.g.} for taking advantage of the M\"obius invariance to fix $z_1$ at $\infty$. We can find a form of these polynomials that is more convenient as follows. Write
 $k_{a_1a_2\ldots a_n}=k_{a_1}+k_{a_2}+\dots+k_{a_n}$,
\begin{align}
g_2(z,k)
&=\sum_{a,b\in A\atop a\ne b}k_a\cdot k_b\,z_a^2
+{1\over 2}\sum_{a,b\in A\atop a\ne b}k_a\cdot k_b\,z_az_b\cr
&=-\sum_{a\in A}k_a^2z_a^2
+{1\over 2}\sum_{a,b\in A\atop a\ne b}k_a\cdot k_bz_az_b={1\over 4}\sum_{a,b\in A\atop a\ne b}k_{ab}^2z_az_b.\label{g2}
\end{align}
\begin{align}
g_3(z,k)&=\sum_{a,b\in A\atop a\ne b}k_a\cdot k_bz_a^3
+\sum_{a,b\in A\atop a\ne b}k_a\cdot k_bz_a^2z_b\cr
&=-\sum_{a\in A}k_a^2z_a^3
+\sum_{a,b\in A\atop a\ne b}k_a\cdot k_bz_a^2z_b={1\over 2}\sum_{a,b\in A\atop a\ne b}k_{ab}^2z_a^2z_b.\label{g3}
\end{align}
\begin{align}
g_2(z,k)\sum_{c\in A}z_c
&={1\over 2}\sum_{a,b\in A\atop a\ne b}k_{ab}^2z_a^2z_b
+{1\over 4}\sum_{a,b,c \in A\atop a, b,c \hbox{ \tiny uneq.}}k_{ab}^2z_az_bz_c\cr
&={1\over 2}\sum_{a,b\in A\atop a\ne b}k_{ab}^2z_a^2z_b
+{1\over 12}\sum_{a,b,c \in A\atop a, b,c \hbox{ \tiny uneq.}}k_{abc}^2z_az_bz_c,
\end{align}
since $k_{abc}^2=k_{ab}^2+k_{bc}^2+k_{ac}^2$, given that $k_b^2=0$ for $b\in A$. Thus,
\be
\sum_{c,a\in A\atop c\ne a}z_cz_a^3f_a(z,k)=
g_2(z,k)\sum_{c\in A}z_c-g_3(z,k)={1\over 12}\sum_{a,b,c \in A\atop a, b,c \hbox{ \tiny uneq.}}k_{abc}^2z_az_bz_c.\label{th3}
\ee
More generally,
\begin{align}
\sum_{a_0,a_2,\ldots,a_m\in A\atop a_i \hbox{ \tiny uneq.}} z_{a_2}&\ldots z_{a_m}z_{a_0}^2f_{a_0} \cr
&=\sum_{a_0,a_1,\ldots,a_m\in A\atop a_i \hbox{ \tiny uneq.}}{k_{a_0}\cdot k_{a_1}z_{a_0}^2z_{a_2}\ldots z_{a_m}\over z_{a_0}-z_{a_1}}
+(m-1)\sum_{a_0,a_1,\ldots,a_{m-1}\in A\atop a_i \hbox{ \tiny uneq.}}{k_{a_0}\cdot k_{a_1}z_{a_0}^2z_{a_1}\ldots z_{a_{m-1}}\over z_{a_0}-z_{a_1}}\cr
&=\sum_{a_0,a_1,\ldots,a_m\in A\atop a_i \hbox{ \tiny uneq.}}k_{a_0}\cdot k_{a_1}z_{a_0}z_{a_2}\ldots z_{a_m}
+{m-1\over 2}\sum_{a_0,a_1,\ldots,a_{m-1}\in A\atop a_i \hbox{ \tiny uneq.}}k_{a_0}\cdot k_{a_1}z_{a_0}z_{a_1}\ldots z_{a_{m-1}}\cr
&=-{m-1\over 2}\sum_{a_0,a_1,\ldots,a_{m-1}\in A\atop a_i \hbox{ \tiny uneq.}}k_{a_0}\cdot k_{a_1}z_{a_0}z_{a_1}\ldots z_{a_{m-1}}.\cr
\end{align}
So
\begin{align}
\sum_{a_0,a_2,\ldots,a_m\in A\atop a_i \hbox{ \tiny uneq.}} z_{a_2}\ldots z_{a_m}z_{a_0}^2f_{a_0}
&=-{1\over m}\sum_{a_1,a_2,\ldots,a_{m}\in A\atop a_i \hbox{ \tiny uneq.}}\sum_{i,j=1\atop i<j}^mk_{a_i}\cdot k_{a_j}z_{a_1}z_{a_2}\ldots z_{a_m}\cr
&=-{1\over 2m}\sum_{a_1,a_2,\ldots,a_{m}\in A\atop a_i \hbox{ \tiny uneq.}}k_{a_1a_2\ldots a_m}^2z_{a_1}z_{a_2}\ldots z_{a_m},\cr
\noalign{\vskip-6pt\hfill which vanishes unless $2\leq m\leq N-2$,\vskip6pt}
&=-{(m-1)!\over 2}\sum_{S\subset A\atop |S|=m}k_S^2z_S,\label{thm}
\end{align}
where $|S|$ denotes the number of elements of $S$, and $k_S, z_S$ are defined by (\ref{kS}).

The homogeneous polynomials 
\be
\tilde h_m=\sum_{S\subset A\atop |S|=m}k_S^2z_S={1\over m!}\sum_{a_1,a_2,\ldots,a_{m}\in A\atop a_i \hbox{ \tiny uneq.}}k_{a_1a_2\ldots a_m}^2z_{a_1}z_{a_2}\ldots z_{a_m},\quad 2\leq m\leq N-2,  \label{defth}
\ee
are quite simply related to the $g_m$ defined by (\ref{defg}) because
\begin{align}
\sum_{a_0,a_2,\ldots,a_m\in A\atop a_i \hbox{ \tiny uneq.}} z_{a_2}\ldots z_{a_m}z_{a_0}^2 f_{a_0}& 
=\sum_{a_2,\ldots,a_m\in A\atop a_i \hbox{ \tiny uneq.}} z_{a_2}\ldots z_{a_m}\sum_{a_0\in A}z_{a_0}^2 f_{a_0}-(m-1)\sum_{a_0,a_3,\ldots,a_m\in A\atop a_i \hbox{ \tiny uneq.}} z_{a_3}\ldots z_{a_m}z_{a_0}^3 f_{a_0}\cr
&=\sum_{r=2}^{m+1}{(-1)^r(m-1)!\over (m-r+1)!}\sum_{a_r,\ldots,a_m\in A\atop a_i \hbox{ \tiny uneq.}} z_{a_r}\ldots z_{a_m}\sum_{a_0\in A}z_{a_0}^r f_{a_0},
\end{align}
so that, since $g_1$ vanishes identically,
\be
\tilde h_m=2\sum_{r=2}^m(-1)^rg_r\Sigma^A_{m-r}, \quad 2\leq m\leq N-2, \label{thgS}
\ee
where $\Sigma_r^A$ is the elementary symmetric function
\be
\Sigma_r^A =\sum_{S\subset A\atop |S|=r}z_S={1\over r!}\sum_{a_{1},\ldots,a_r\in A\atop a_i \hbox{ \tiny uneq.}} z_{a_{1}}\ldots z_{a_r}.\label{symfn}
\ee
Since the equations $f_a=0, a\in A,$ are equivalent to the equations $g_m=0,\; 2\leq m\leq N-2$,  they are also equivalent to the equations, $\tilde h_m=0, \;2\leq m\leq N-2$.
$\tilde h_m$ is a homogeneous polynomial of degree $m$ with the special property that, although it is of degree $m$ in the $z_a$, it is linear in each one of them taken individually, {\it i.e.} the monomials that it comprises are square-free. This considerably simplifies various calculations and constructions. 

The coefficients, $k_{a_1a_2\ldots a_m}^2$, in the $\tilde h_m$ are all determined in terms of the $\half N(N-1)$ coefficients $k_{ab}^2=2k_a\cdot k_b$,
\be
k_U^2=\sum_{S\subset U\atop |S|=2}k_S^2.\label{kU2}
\ee
 Of course, these are not independent but must satisfy the constraints of momentum conservation. This leaves $\half N(N-3)$ independent $k_a\cdot k_b$, provided that $N\leq D+1$, so that the constraints of the dimension $D$ of space-time do not enter. If $N>D+1$, the number of independent $k_a\cdot k_b$ is reduced to $N(D-1)-\half D(D+1)$.
 
 {\it An Alternative Derivation. } We can give a more elegant and succinct, if less direct, derivation of the polynomial form of the scattering equations as follows. As we noted in section \ref{Introduction}, the scattering equations (\ref{SE}) are equivalent to $p(z)^2$, as given by 
(\ref{p2}), vanishing everywhere as a function of $z$ for $z_a, a\in A,$ satisfying these equations, and this statement is equivalent to the vanishing of the polynomial of degree $N-2$, 
\be
F(z)=2p(z)^2\prod_{a\in A} (z-z_a) =\sum_{S\subset A\atop |S|=2}k^2_S \prod_{b\in\bar S}(z-z_b),\label{defF}
\ee
where $\bar S=\{a\in A:a\notin S\}$. Now, 
\be
F(z)=\sum_{m=0}^{N-2}z^{N-m-2}\sum_{U\subset A\atop |U|=m}z_U\sum_{S\subset \bar U\atop |S|=2}k^2_S.
\ee
From (\ref{kU2}), 
\be
\sum_{S\subset \bar U\atop |S|=2}k^2_S=k^2_{\bar U}=k^2_U
\ee
so that the coefficients of $z^{N-2}$ and $z^{N-3}$ vanish and
\be
F(z)=\sum_{m=2}^{N-2}z^{N-m-2}\tilde h_m.
\ee
establishing the equivalence of the polynomial form (\ref{PSE}) to the original scattering equations (\ref{SE}).

{\it Partially Fixing the M\"obius Invariance. } Because the set is equivalent to the equations $f_a=0, a\in A$, the M\"obius group acts on the solutions of the set of equations $\tilde h_m=0, 2\leq m\leq N-2$ (which define an algebraic variety in $\Cop^N$). As usual, it is convenient to fix this invariance, at least partially. We take $z_1\rightarrow \infty, z_N\rightarrow0.$ This is facilitated by the fact that $\tilde h_m$ is linear in each of $z_1,z_N$ separately. Write $A'=\{a\in A: a\ne 1, N\}$ and
\be
h_m=\lim_{z_1\rightarrow\infty}{\tilde h_{m+1}\over z_1}=\sum_{S\subset A'\atop |S|=m}k^2_{S_1}z_S={1\over m!}\sum_{a_1,a_2,\ldots,a_{m}\in A'\atop a_i \hbox{ \tiny uneq.}}\sigma_{a_1a_2\ldots a_m}z_{a_1}z_{a_2}\ldots z_{a_m},\label{defh}
\ee
where
\be
S_1= S\cup\{1\},\qquad \sigma_{a_1a_2\ldots a_m}=k_{1a_1a_2\ldots a_m}^2.
\ee
$h_m$ is an homogeneous polynomial of degree $m$ in the variables $z_2, \ldots, z_{N-1},$ linear in each of them separately. Fixing $z_1=\infty, z_N=0$ breaks the M\"obius group down to scalings $z_a\mapsto \lambda z_a, a\in A'.$
At $z_1=\infty, z_N=0$, we can replace the equations $\tilde h_m=0, 2\leq m\leq N-2$ by $ h_m=0, 1\leq m\leq N-3.$ These equations define a (presumably) zero-dimensional projective variety in $\Cop\Pop^{N-3}$, a set consisting typically of 
\be
\prod_{m=1}^{N-3}\deg h_m=(N-3)! 
\ee
points by B\'ezout's Theorem.

{\it Factorization.} To see how the factorization, evident in  (\ref{SES}) and (\ref{SES1}), when $z_a\rightarrow z_0,\;a\in S,$ implying $k_S^2=0$, manifests itself in terms of the polynomial equations, consider the equations $\tilde h_m=0, 2\leq m\leq N-2,$ 
without specializing to $z_1=\infty, z_N=0$, and use the translation invariance that then holds to take $z_0=0,$ without loss of generality. Let $n=|\bar S|$, where again $\bar S=\{a\in A:a\notin S\}$, and $z_a=\epsilon x_a+\O(\epsilon^2), a\in S$. Then 
\be
\tilde h_m=\tilde h_m^{\bar S0}+\O(\epsilon),\qquad 1<m< n,
\ee
where
\be
\tilde h_m^{\bar S0}=\sum_{U\subset \bar S\atop |U|=m}k_{U}^2z_U,
\ee
$\tilde h_n$ vanishes because $k_S^2=0$, and, for $n<m<N$,
\be
\tilde h_m=\epsilon^{n-m}\tilde h_{m-n}^{S\infty}z_{\bar S}+\O(\epsilon^{m-n+1}),
\ee
where
\be
\tilde h_r^{S\infty}={1\over r!}\sum_{U\subset S\atop |U|=r}(k_{\bar S}+k_{U})^2x_{U},\qquad 1\leq r<N-n.
\ee
$\tilde h_m^{\bar S0}$ is the appropriate form of  $\tilde h_m$ for the polynomial scattering equations for $n+1$ particles with variables $z_a, a\in\bar S$ and $0$, associated with momenta $k_a,a\in\bar S,$ and $k_S$, while 
$\tilde h_m^{S\infty}$ is the appropriate form of  $ h_m$ for the polynomial scattering equations for $N-n+1$ particles with variables $z_b, b\in\ S,$ and $\infty$, associated with momenta $k_b,b\in S,$ and $k_{\bar S}$, thus demonstrating that the polynomial equations factorize as they should. 

\section{Solutions to the Scattering Equations}
\label{Solutions}

{\it A Special Symmetric Configuration.}  For a given number of particles, $N$, if the dimension of space-time, $D\geq N-1$, we choose momenta,  $k_a$, so that the coefficients in $h_m$, for a given $m$, are all equal, by arranging that 
$k_a\cdot k_b=\mu, k_1\cdot k_a =-\half (N-3)\mu,$ for $a,b\in A'.$ [For example, in $\Rop^{1,N-2}$, take $k_1^0=k_N^0=\alpha,k_1^j=-k_N^j=\beta, k_a^0=\gamma,k_a^j=\delta, j\ne a-1, k_a^{a-1}=-(N-3)\delta, 2\leq a\leq N-1,$ where
$4\alpha^2=4(N-2)\beta^2=(N-2)^2\gamma^2=(N-3)(N-2)^3\delta^2$, so that $\mu=(N-2)^2\delta^2$. The configuration describes massless particles with momenta $k_1,k_N$ annihilating in their center of mass frame to produce symmetrically $N-2$ massless particles with momenta $k_a, 2\leq a\leq N-1$, the spatial components of whose momenta correspond to the vertices of a regular $(N-3)$-simplex.]

If its coefficients $k_{1a_1\ldots a_m}$ are all equal, $h_m$ is proportional to the elementary symmetric function, $\Sigma_m^{A'},$ of (\ref{symfn}). The conditions $\Sigma_m^{A'}=0, 1\leq m\leq N-3,$ imply that 
\be
\prod_{b=2}^{N-1}(z-z_b)=z^{N-2}-\lambda^{N-2}, \qquad\hbox{where}\quad \lambda^{N-2}=\prod_{b=2}^{N-1}z_a.
\ee
Thus each $z=z_a$, $2\leq a\leq N-1$, satisfies the equation $z^{N-2}=\lambda^{N-2}$ and, taking $\lambda =1$, because only ratios of the $z_a$ matter, the solutions of the scattering equations for these momenta are
\be
z_a=\omega^{\rho(a-1)},\qquad \rho\in{\frak S}_{N-2},\qquad 2\leq a\leq N-1,
\ee
and $\omega$ is a complex $(N-2)$-th root of unity. The ratios $z_a/z_2$, $3\leq a\leq N-1$, say, are thus given by the $N-3$ distinct complex $(N-2)$-th roots of unity, taken in some order, giving the $(N-3)!$ possible solutions. For this particular symmetric solution there is a symmetry under permuting the coordinates, which takes one solution into another, which is not present in general. 

For further discussion of special configurations see \cite{K}, where Kalousios discusses  kinematical configurations in which the solutions to the scattering equations can be identified with the zeros of Jacobi  polynomials, and \cite{CHY2}.

{\it Solutions for $N=4$ and $N=5$.}  
For $N=4$, we have a simple linear equation determining $z_3/z_2$,
\be
h_1=\sigma_2z_2+\sigma_3z_3=0,\quad z_3/z_2=-\sigma_2/\sigma_3=-k_1\cdot k_2/k_1\cdot k_3.
\ee

For $N=5$, writing $(x,y,z)=(z_2,z_3,z_4)$, the equations
\begin{align}
h_1&=\sigma_2x+\sigma_3y+\sigma_4z=0\cr
h_2&=\sigma_{23}xy+\sigma_{24}xz+\sigma_{34}yz=0
\end{align}
yield a quadratic equation for $z/y$ by elimination of $x$. This can be conveniently achieved by considering
\be
h_{12}^{(x)}=\bd h_1&h_2\cr h_1^{x}&h_2^{x}\ed, 
\ee
where
\be
 h^x_m={\partial h_m\over\partial x},\quad\hbox{and, more generally, write}\quad h_m^{a_1\ldots a_r}= {\partial^r h_m\over \partial z_{a_1}\ldots\partial z_{a_r}}.
\ee
Then, because $h_m$ is linear in each $z_i$ taken separately, $h_m^{a_1\ldots a_r}=0$ if $a_i=a_j$ for any $i\ne j$. Additionally, $h_m^{a_1\ldots a_r}=0$ if $r>m$. It follows immediately that $h_{12}^{(x)}$ is independent of $x$ because 
\be
{\partial h_{12}^{(x)}\over \partial x}=\bd h_1^{x}&h_2^{x}\cr h_1^{x}&h_2^{x}\ed+\bd h_1&h_2\cr h_1^{xx}&h_2^{xx}\ed=0.\label{h12x}
\ee
$h^{(x)}_{12}$ is quadratic in $y$ and $z$ (as can be seen, {\it e.g.}, by noting, using the approach of  (\ref{h12x}), that any third derivative of $h^{(x)}_{12}$ with respect to $y$ and $z$ vanishes) and $h^{(x)}_{12}=0$  when $h_1=h_2=0$, so that it provides the required quadratic equation for $y/z$. Given a solution to this equation, $x/z$ is determined uniquely from $h_1=0$.

{\it Solution for $N=6$.}  This approach can be extended to $N=6$ without much difficulty, where we expect a sextic equation to determine the ratios $z_a/z_b$. Write $(x,y,z,u)=(z_2,z_3,z_4,z_5)$ and 
\be
h_{123}^{(x|y)}=\bd h_1&h_2&h_3\cr h_1^{x}&h_2^{x}&h_3^{x}\cr h_1^{xy}&h_2^{xy}&h_3^{xy}\ed. 
\ee
Then the technique of (\ref{h12x}) shows that $\partial_xh_{123}^{(x|y)}=\partial_y^2h_{123}^{(x|y)}=0$, implying that $h_{123}^{(x|y)}$ is independent of $x$ and linear in $y$; similarly it is cubic in $z$ and $u$, suggesting that the required sextic might involve the product of two such expressions, $h_{123}^{(x|y)}h_{123}^{(y|x)}$. Indeed, 
\be
h_{123}^{(xy)}=\bd h_1&h_2&h_3\cr h_1^{x}&h_2^{x}&h_3^{x}\cr h_1^{xy}&h_2^{xy}&h_3^{xy}\ed\bd h_1&h_2&h_3\cr h_1^{y}&h_2^{y}&h_3^{y}\cr h_1^{xy}&h_2^{xy}&h_3^{xy}\ed
-\bd h_1&h_2&h_3\cr h_1^{x}&h_2^{x}&h_3^{x}\cr h_1^{y}&h_2^{y}&h_3^{y}\ed\bd h_1^{x}&h_2^{x}&h_3^{x}\cr h_1^{y}&h_2^{y}&h_3^{y}\cr h_1^{xy}&h_2^{xy}&h_3^{xy}\ed,\label{h123xy}
\ee
is indeed independent of both $x$ and $y$ because $\partial_x h_{123}^{(xy)}=\partial_yh_{123}^{(xy)}=0$, and it vanishes when $h_1=h_2=h_3=0$; so $h_{123}^{(xy)}=0$, provides the required sextic in $z, u$.
Given one of the solutions for $z/u$ to $h_{123}^{(xy)}=0$, $x/u$ and $y/u$ are uniquely determined by $h_{123}^{(y|x)}=0$ and $h_{123}^{(x|y)}=0$, respectively.

The definition (\ref{h123xy}) can be rewritten as a single determinant,
\be
h_{123}^{(xy)}=-\bd h_1&h_2&h_3&0&0&0\cr h_1^{x}&h_2^{x}&h_3^{x}&0&0&0\cr h_1^{y}&h_2^{y}&h_3^{y}&h_1&h_2&h_3\cr h_1^{xy}&h_2^{xy}&h_3^{xy}&h_1^{x}&h_2^{x}&h_3^{x}\cr0&0&0&h_1^{y}&h_2^{y}&h_3^{y}\cr0&0&0&h_1^{xy}&h_2^{xy}&h_3^{xy}\cr\ed,
\ee
which is a form that can be obtained from classical elimination theory \cite{GS}.

\section{Amplitudes in Terms of Polynomial Constraints}
\label{Amplitudes}

We want to rewrite (\ref{amp}) in terms of $h_m, 1\leq m\leq N-3$, rather than $f_a, a\in A, a\ne i,j,k.$ We can invert (\ref{defg}) to give
\be
f_a(z,k)=\sum_{m=2}^{N-2}(Z^{-1})_{am}g_m(z,k),
\ee
where $Z_{ma}=z_a^{m+1}, a\in A, -1\leq m\leq N-2,$ noting that $g_{-1}=g_0=g_1=0$. Since $Z^{-1}=(\det Z)^{-1} \hbox{adj} Z$, the Jacobian of $\{f_a:a\in A, a\ne i,j,k\}$ with 
respect to $\{g_m: 2\leq m\leq N-2\}$ is given by
\begin{align}
\det\left({\partial f_a\over\partial g_m}\right)_{a\ne i,j,k\atop 2\leq m\leq N-2}&={1\over (\det Z)^{N-3}}\det\left( \hbox{adj} Z\right)_{a\ne i,j,k\atop 2\leq m\leq N-2}
={1\over \det Z}\det\left( Z\right)_{a= i,j,k\atop m=-1,0,1}\cr
&=(z_i-z_j)(z_i-z_k)(z_j-z_k) \prod_{a<b}(z_a-z_b)^{-1},\\
\det\left({\partial \tilde h_m\over\partial g_n}\right)_{2\leq m,n\leq N-2}&=(-1)^{\half N(N+1)}2^{N-3}.
\end{align}
Thus, up to a sign and factors of 2,
\be
\A_N=\oint_\O\Psi_N(z,k)\prod_{m=2 }^{N-2}{1\over \tilde h_m(z,k)}\prod_{a<b} (z_a-z_b)\prod_{a\in A}{dz_a\over (z_a-z_{a+1})^2}\bigg/d\omega.\label{amp1}
\ee
Taking $z_1\rightarrow\infty, z_2$ fixed, $ z_N\rightarrow 0$,
\be
\A_N=\oint_\O\Psi_N(z,k){z_2\over z_{N-1}}\prod_{m=1 }^{N-3}{1\over h_m(z,k)}\prod_{2\leq a<b\leq N-1} (z_a-z_b)\prod_{a=2}^{N-2}{z_adz_{a+1}\over (z_a-z_{a+1})^2}.\label{amp2}
\ee

\section{M\"obius Transformations} 
\label{Moebius}

{\it M\"obius Invariance of the Integrand for the Amplitude.} It is straightforward to demonstrate directly the M\"obius invariance of (\ref{amp1}). The M\"obius group is generated by translations, scaling and inversion, and,
in this instance, demonstrating invariance under translation, $z_a\mapsto z_a+\epsilon,$ involves a little more than the others. We have
\begin{align}
\tilde h_m(z-\epsilon)&=\sum_{S\subset A\atop |S|=m} k_S^2 \prod_{a\in S}(z_a-\epsilon)\\
&=\sum_{S\subset A\atop |S|=m} k_S^2\sum_{r=0}^m (-\epsilon)^r\sum_{U\subset S\atop |U|=m-r}z_U\cr
&=\sum_{r=0}^m (-\epsilon)^r\sum_{U\subset A\atop |U|=m-r}z_U\sum_{U\subset S\subset A\atop |S|=m} k_S^2.
\end{align}
Given $U$ with $|U|=n,$
\begin{align}
\sum_{U\subset V\subset A\atop |V|=n+1} k_V^2&=\sum_{b\in\bar U} (k_U+k_b)^2\cr
&=(N-n)k_U^2+2\sum_{b\in\bar U} k_U\cdot k_b\cr
&=(N-n-2)k_U^2,\qquad\hbox{as }\; k_U=-k_{\bar U}.
\end{align}
Thus
\be
\sum_{U\subset S\subset A\atop |S|=m} k_S^2={ (N-m+r-2)!\over r!(N-m-2)!}k_U^2,
\ee
and so
\be
\tilde h_m(z_1-\epsilon,\ldots,z_N-\epsilon)=\sum_{r=0}^{m-2} { (N-m+r-2)!\over r!(N-m-2)!}\,(-\epsilon)^r\,\tilde h_{m-r}(z_1,\ldots,z_N),\label{trans}
\ee
as $\tilde h_0(z)=\tilde h_1(z)=0$,
from which it follows that (\ref{amp1}) is unchanged under translations. 
Under scaling and inversion, respectively,
\begin{align}
 \tilde h_m (\lambda z_1,\dots,\lambda z_N)&=\lambda^m \tilde h_m ( z_1,\dots, z_N),\label{scale}\\
\tilde h_m  (1/ z_1,\dots,1/ z_N)&=\tilde h_{N-m}(z_1,\ldots,z_N)/z_A\label{inv}
\end{align}
and it may easily be checked that (\ref{amp1})  is invariant under these as well, so that it is invariant under the full M\"obius group.

{\it Representations of the M\"obius Algebra.} Combining translation, $T_\epsilon$, (\ref{trans}), with inversion, $I$, (\ref{inv}), gives the special conformal transformation, $S_\epsilon=IT_{-\epsilon} I$, 
\be
\tilde h_m\left({z_1\over 1+\epsilon z_1},\ldots,{z_N\over 1+\epsilon z_N}\right)=\sum_{r=0}^{N-m-2} { (m+r-2)!\over r!(m-2)!}\,\epsilon^r\,\tilde h_{m+r}(z_1,\ldots,z_N)\prod_{a\in A}{1\over 1+\epsilon z_a}.\label{spec}
\ee

If $L_{-1}$ denotes the generator of translations (\ref{trans}),
\be
L_{-1}=-\sum_{a\in A}{\partial\over \partial z_a},\qquad L_{-1}\,\tilde h_m=- (N-m-1)\tilde h_{m-1}\,,\label{L-1}
\ee
the generator of special conformal transformations (\ref{spec}),
\be
L_{1}=-\sum_{a\in A}z_a^2{\partial\over \partial z_a}+\Sigma_1^A,\qquad L_{1}\,\tilde h_m=(m-1)\tilde h_{m+1}.\label{L1}
\ee
With these definitions, the appropriate generator of scale transformations is 
\be
L_{0}=-\sum_{a\in A}z_a{\partial\over \partial z_a}+{N\over 2},\qquad L_{0}\,\tilde h_m =(\half N-m)\tilde h_{m}\, \label{L0}
\ee
so that
\be
[L_1,L_{-1}]=2L_0,\qquad [L_0,L_{\pm 1}]=\mp L_{\pm 1}.\label{alg}
\ee

The $\tilde h_m, \,2\leq m\leq N-2,$ form a basis for an $(N-3)$-dimensional representation of the M\"obius algebra, {\it i.e.} a representation of `M\"obius spin' $\half N-2$. For this representation the quadratic Casimir,
\be
L^2\equiv L_0^2-\half L_1L_{-1}-\half L_{-1}L_1,
\ee
takes the value $(\half N-2)(\half N-1)$.

More generally, the ring, $\R^N$, of polynomials in $z_1, z_2, \ldots, z_N$, provides a graded infinite-dimensional representation space for (\ref{alg}). This is the $N$-fold tensor product of the representation 
\be
L_{-1}=-{d\over dz},\qquad L_0=-z{d\over dz}+{1\over 2},\qquad L_1=-z^2{d\over dz} +z,\label{Lz}
\ee
acting on the ring of single-variable polynomials, $\R^1$, in $z$. It is easy to see that the only invariant subspace of $\R^1$ for (\ref{Lz}) is the two-dimensional subspace, $\F^1$,  consisting of  linear polynomials, {\it i.e.} that spanned by $\{1,z\}$, which carries a representation of `M\"obius spin'  ${1\over 2}$, for which $L^2={3\over 4}$. [The quotient space $\R^1/\F^1$ provides an irreducible infinite-dimensional representation of the M\"obius algebra with `M\"obius spin' ${1\over 2}$ and $L_0\leq -{3\over 2}$.\big] Correspondingly, $\R^N$ has a M\"obius invariant subspace $\F^N$, which is the tensor product of $N$ copies of $\F^1$, one for each $z_a, a\in A$, and consists of polynomials that are linear in each of the $z_a$ taken separately. $\F^N$ has dimension $2^N$ and has a basis consisting of the square-free monomials, $\{z_S: S\subset A\}$. 

It is straightforward to show that, if the polynomial $\varphi(z)\in\R^N$ involves a monomial term with a factor $z_a^n$, with $n\geq 2$, then $(L_1)^M\varphi$ involves a monomial term with a factor $z_a^{M+n}$, $M\geq 0$, implying that $\varphi$ can not be an element of a finite-dimensional M\"obius invariant subspace of $\R^N$. Thus $\F^N$ is the largest finite-dimensional M\"obius invariant subspace. It decomposes into irreducible subspaces of `M\"obius spin'  $\half N-n$, $n$ an integer, $0\leq n\leq {1\over 2}N$, of dimension $N-2n +1$, with the eigenvalues of $L_0$ being $\half N- m,$ where $n\leq m\leq N-n$. 

{\it Highest Weight Polynomials.} Such irreducible subspaces,  of `M\"obius spin'  $\half N-n$,  are generated by the repeated action of $L_1$ from highest weight polynomials $\varphi$ of degree $n$ satisfying $L_{-1}\varphi=0$, {\it i.e.} translation invariant polynomials in $\F^N$. The highest weight polynomial in an irreducible subspace is the one with the largest eigenvalue of $L_0$, or, equivalently, the lowest degree. If $\F^N_m$ denotes the subspace of $\F_N$ consisting of polynomials of degree $m$, so that $\F^N_m$ has a basis $\{z_S:S\subset A, |S|=m\}$,  and $\H^N_n=\{\varphi\in\F^N_n:L_{-1}\varphi=0\}$, the subspace of $\F^N_m$ consisting of highest weight polynomials, 
\be
\dim\F^N_m={N!\over m!(N-m)!};\qquad \dim\H^N_n = \dim\F^N_n - \dim\F^N_{n-1} ={N!(N-2n+1)\over n!(N-n+1)!},\label{dimFH}
\ee
where $0\leq m\leq N$, $0\leq n\leq \half N$. Thus,
\be
\dim\H^N_0 = 1;\qquad \dim\H^N_1 = N-1;\qquad \dim\H^N_2 = {N(N-3)\over 2}.
\ee

For $\varphi\in\F^N_n$, write 
\be
\varphi=\sum_{S\subset A\atop |S|=n}\varphi_Sz_S,\qquad \varphi_S=\varphi_{i_1\ldots i_n}, \;\;\hbox{ for }\;\; S=\{i_1,\ldots, i_n\}.\label{varphi}
\ee
[Here $\varphi_{i_1\ldots i_n}$ is symmetric in $i_1,\ldots, i_n$ and vanishes if any two indices are equal.]
Then $\varphi\in\H^N_n$, {\it i.e.} $L_{-1}\varphi=0$, if and only if
\be
\sum_{S\subset A, S\ni a\atop |S|=n}\varphi_S=0, \;\hbox{ for each }\; a\in A,\;\; \hbox{or, equivalently, }\;\;\sum_{i_n\in A}\varphi_{i_1\ldots i_n}=0.\label{phicon}
\ee
We can verify directly that the dimension of the space of tensors $\varphi_{i_1\ldots i_n}$ satisfying these conditions is $ \dim\H^N_n$ as given by (\ref{dimFH}). 
The action of $L_1$ on $\varphi\in\H^N_n$, defined by (\ref{varphi}), is given by
\be
L_1\varphi=\sum_{S\subset A\atop |S|=n}\varphi_S\sum_{a\in \bar S}z_az_S=\sum_{U\subset A\atop |U|=n+1}\varphi_Uz_U,
\ee
where, in general, we define
\be
\varphi_U=\sum_{S\subset U\atop |S|=n}\varphi_S.\label{phiU}
\ee
Then, if $m\geq n,$ and
\be
\varphi_m=\sum_{U\subset A\atop |U|=m}\varphi_Uz_U,\label{defphim}
\ee
we have
\be
 L_1\varphi_m=\sum_{U\subset A\atop |U|=m}\sum_{a\in \bar U}z_az_U\sum_{S\subset U\atop |S|=n}.
\varphi_S=(m-n+1)\varphi_{m+1}
\ee
So $L^r_1\varphi=r!\varphi_{n+r}$ and, using the algebra (\ref{alg}) and  $L_{-1}\varphi=0$, we can calculate $L_{-1}L^r_1\varphi$ and deduce that 
\be
L_1\varphi_m=(m-n+1)\varphi_{m+1},\quad L_0\varphi_m=(\half N-m)\varphi_m,\quad L_{-1}\varphi_m=-(N-m-n+1)\varphi_{m-1},
\ee
where $\varphi_{n-1}=0$, and, as we see below, $\varphi_{N-n+1}=0$, and 
\be
L^2\varphi_m=(\half N-n)(\half N-n+1)\varphi_m,
\ee
in agreement with the `M\"obius spin'  of the multiplet with highest weight polynomial $\varphi\equiv\varphi_n$ being $\half N-n$, and with (\ref{L-1})--(\ref{L0}) for $n=2$. It follows from (\ref{phicon}) that 
\begin{align}
\varphi_U&={1\over n!}\sum_{a_1,\ldots,a_n\in U\atop}\varphi_{a_1\ldots a_n}=-{1\over n!}\sum_{a_1,\ldots,a_{n-1}\in U\atop a_{n}\in\bar U}\varphi_{a_1\ldots a_n}\cr
&={(-1)^n\over n!}\sum_{a_1,\ldots,a_n\in \bar U\atop}\varphi_{a_1\ldots a_n}=(-1)^n\varphi_{\bar U}.
\end{align}
It follows that $I\varphi_m=(-1)^n\varphi_{N-m}$, from which it follows that $\varphi_m=0$ if $m>N-n,$ consistent with the fact that the $N-2n+1$ polynomials 
$\varphi_n, \varphi_{n+1},\ldots, \varphi_{N-n}$ are a basis  for the invariant subspace generated from the highest weight polynomial $\varphi\equiv\varphi_n$.

A basis for $\F^N_0$ is provided by $\varphi\equiv\varphi_0=1$, and then $\varphi_m=\Sigma^A_m$, $0\leq m\leq N$, the elementary symmetric polynomial defined by (\ref{symfn}). An element of $\F^N_1$ is of the form $\varphi\equiv\varphi_1=\sum_{a\in A}\lambda_az_a,$ where $\sum_{a\in A}\lambda_a=0,$ and then
\be
\varphi_m = \sum_{a\in A}\lambda_a z_a\Sigma_{m-1}^{A_a},\qquad\hbox{where }\; A_a=\{b\in A: b\ne a\}.
\ee

{\it Uniqueness.} If $\varphi\in\H^N_n$, the invariant linear subspace generated from $\varphi$ has dimension $N-2n+1$, and so it is the case $n=2$ that is of particular interest to us because it is only in this case that the conditions $\varphi_m=0$ provide the right number of constraints to determine a finite number of points in $\Cop\Pop^{N-1}$, up to M\"obius transformations. Further, from (\ref{dimFH}), $\dim\H^N_2=\half N(N-3),$ which, if the dimension of space-time is sufficiently high, is the number of independent degrees of freedom, $k_a\cdot k_b$, of light-like momenta $k_a, a\in A$, satisfying momentum conservation (up to Lorentz transformation), with $\varphi_U$ determined by 
\be
\varphi_U=k_U^2 \label{phiUk}
\ee
as in (\ref{kU2}). This implies that a M\"obius invariant set of equations in $\Cop\Pop^{N-1}$ determining a finite set of points, up to M\"obius transformation, has to have the form of the polynomial scattering equations (\ref{PSE}), at least if the M\"obius transformations are realized as in (\ref{L-1})--(\ref{L0}).

To see more directly that, if $\varphi\in\H^N_2$, then $\varphi_m$ is of the form  (\ref{PSE}), take vectors, $k_a, a\in A$, 
 summing to zero, with $\{k_1,\ldots, k_{N-1}\}$ linearly independent, define a scalar product by
\be
\langle k_a,k_b\rangle=\langle k_b,k_a\rangle=\lambda_{ab},\quad 1\leq a<b\leq N-1;\quad \langle k_a,k_a\rangle=0,\quad 1\leq a\leq N-1.\label{kakb}
\ee
Then, taking $\varphi_{a,b}=(k_a+k_b)^2$, $\varphi_U$, as defined by (\ref{phiU}), is given by (\ref{phiUk}) and $\varphi_m=\tilde h_m,$ defined as in (\ref{defth}).

\section{Massive Particles}
\label{Massive}

In the discussion of massless particles, the order of the particles is not relevant. To discuss massive particles \cite{DG1}, we need to select a cyclic ordering  on $A$, which we take to be 
the ordering $1,2,\ldots, N, 1$. We replace the definition of $f_a$ in (\ref{SE}) by 
\be
f_a(z,k)=\sum_{b\in A\atop b\ne a}{k_a\cdot k_b\over z_a-z_b}+{\mu^2\over 2(z_a-z_{a-1})}+{\mu^2\over 2(z_a-z_{a+1})},\quad a\in A.\label{MSE}
\ee
Using (\ref{MSE}) in (\ref{defg}) to define $g_m$, it is again the case that M\"obius invariance of this system of equations implies that $g_{-1},g_0,g_1$ all vanish 
identically, provided that $k_a^2=\mu^2$ for each $a\in A$.
\begin{align}
g_m(z,k)
&={1\over 2}\sum_{a,b\in A\atop a\ne b}k_a\cdot k_b{z_a^{m+1}-z_b^{m+1}\over z_a-z_b}+{\mu^2\over 2}\sum_{a\in A}{z_{a+1}^{m+1}-z_{a}^{m+1}\over z_{a+1}-z_{a}}\cr
&={1\over 2}\sum_{a,b\in A\atop a\ne b}k_a\cdot k_b\sum_{r=0}^{m}z_a^rz_b^{m-r}+{\mu^2\over 2}\sum_{a\in A}\sum_{r=0}^{m}z_{a+1}^rz_a^{m-r} .\label{gdefM}
\end{align}

Then, following section \ref{Polynomial}, we obtain, as in (\ref{g2}),
\be
g_2(z,k)=
{1\over 2}\sum_{a,b\in A\atop a< b}(k_{ab}^2-n_{ab}\mu^2)z_az_b,\label{Mg2}
\ee
where $n_{a,a\pm 1}=1,\; n_{ab}=2,\; b\ne a\pm 1$,  provided that $k_a^2=\mu^2, \;a\in A$, and, as in (\ref{g3}),
\be
g_3(z,k)={1\over 2}\sum_{a,b\in A\atop a\ne b}(k_{ab}^2-n_{ab}\mu^2)z_a^2z_b;\label{Mg3}
\ee
\be
\sum_{c,a\in A\atop c\ne a}z_cz_a^3f_a(z,k)=
g_2(z,k)\sum_{c\in A}z_c-g_3(z,k)={1\over 12}\sum_{a,b,c \in A\atop a, b,c \hbox{ \tiny uneq.}}(k_{abc}^2-n_{abc}\mu^2)z_az_bz_c;
\ee
\begin{align}
g_2(z,k)\sum_{c\in A}z_c
&={1\over 2}\sum_{a,b\in A\atop a\ne b}(k_{ab}^2-n_{ab}\mu^2)z_a^2z_b
+{1\over 4}\sum_{a,b,c \in A\atop a, b,c \hbox{ \tiny uneq.}}(k_{ab}^2-n_{ab}\mu^2)z_az_bz_c\cr
&={1\over 2}\sum_{a,b\in A\atop a\ne b}(k_{ab}^2-n_{ab}\mu^2)z_a^2z_b
+{1\over 12}\sum_{a,b,c \in A\atop a, b,c \hbox{ \tiny uneq.}}(k_{abc}^2-n_{abc}\mu^2)z_az_bz_c,
\end{align}
where $n_{abc}=n_{ab}+n_{ac}+n_{bc}-3$, provided that $k_b^2=\mu^2$ for $b\in A$. 
[$n^2_{abc}=3$, if none of $a,b,c$ are adjacent, $n_{abc}=2$, if just two them are, and $n_{abc}=1$, if $a,b,c$ are sequential (in some order).] Thus, as in (\ref{th3}),
\be
\sum_{c,a\in A\atop c\ne a}z_cz_a^3f_a(z,k)=
g_2(z,k)\sum_{c\in A}z_c-g_3(z,k)={1\over 12}\sum_{a,b,c \in A\atop a, b,c \hbox{ \tiny uneq.}}(k_{abc}^2-n_{abc}\mu^2)z_az_bz_c.
\ee
For $S\subset A$, let $n_S$ denote the number of disjoint strings of adjacent elements of $S$ with the given cyclic ordering on $A$.   Then $1\leq n_S\leq |S|$, with $n_S=1$ if and only if
the elements of $S$ are sequential. Then, as in (\ref{thm}),
\be
\sum_{a_0,a_2,\ldots,a_m\in A\atop a_i \hbox{ \tiny uneq.}} z_{a_2}\ldots z_{a_m}z_{a_0}^2f_{a_0}
=-{(m-1)!\over 2}\sum_{S\subset A\atop |S|=m}(k_S^2-n_S\mu^2)z_S.\label{Mthm}
\ee
We can deduce (\ref{Mg2})--(\ref{Mthm}) directly from section \ref{Polynomial} by introducing a `fictitious' auxiliary space, orthogonal to space-time, and space-like vectors $\kappa_a, a\in A$, with $\kappa_a^2=-1, 
\kappa_a\cdot \kappa_{a+1}=\half, \kappa_a\cdot \kappa_{b}=0, |a-b|>1, a,b\in A$. [The $\kappa_a$, $a\in A$, span an $(N-1)$-dimensional space.] Write $\tilde k_a=k_a+\mu\kappa_a$; then
\be
\tilde k_a^2=k_a^2-\mu^2,\quad a\in A, \qquad\qquad \tilde k_S^2=k_S^2-n_S\mu^2,\quad S\subset A.
\ee
Thus, the massless condition $\tilde k^2=0$ is equivalent to the mass-shell condition $k^2=\mu^2$, and substituting $\tilde k_a$ for $k_a$ in (\ref{SE}) yields (\ref{MSE}) and similarly 
(\ref{Mg2}), (\ref{Mg3}) and (\ref{Mthm}) follow from  (\ref{g2}), (\ref{g3}) and (\ref{thm}), respectively. 

The equations $f_a=0, a\in A,$ are equivalent to the equations $\tilde h_m=0, 2\leq m\leq N-2$,  where 
\be
\tilde h_m=\sum_{S\subset A\atop |S|=m}(k^2_{S}-n_S\mu^2)z_S.
\ee
The $\tilde h_m$ are  related to the $g_m$, defined by (\ref{defg}) with $f_a$ given by (\ref{MSE}), by (\ref{thgS}). The M\"obius invariance of this system can be demonstrated directly as in section \ref{Amplitudes}.

As in section \ref{Polynomial}, it is convenient to fix the M\"obius invariance partially by letting $z_1\rightarrow \infty$ and $ z_N\rightarrow0.$ Again, writing $A'=\{a\in A: a\ne 1, N\}$ and $S_1= S\cup\{1\},$
\be
h_m=\lim_{z_1\rightarrow\infty}{\tilde h_{m+1}\over z_1}=\sum_{S\subset A'\atop |S|=m}(k^2_{S_1}-n_S\mu^2)z_S,\label{defhM}
\ee
we can  replace the equations $\tilde h_m=0, 2\leq m\leq N-2$ by $ h_m=0, 1\leq m\leq N-3.$

\section{Four-Dimensional Space-Time}
\label{FourDim}

Fairlie and Roberts \cite{FR}, who were concerned with four-dimensional space-time, noted that a particular solution of the scattering equations (\ref{SE}) is given by 
\be
z_a={k^3_a-k^0_a\over k^2_a+ik^1_a},\qquad 1\leq a\leq N,
\ee
for $D=4$ and for all $N$. The demonstration of this can conveniently be expressed using twistor variables, $\pi_a, \bpi_a,$ defined by
\be
k_a^\mu\sigma_\mu=\bm k_a^0-k_a^3&-ik_a^1+k_a^2\cr -ik_a^1-k_a^2&k_a^0+k_a^3 \ema 
=\pi_a\bpi_a^T,\qquad \pi_a=\bm\pi_a^1\cr\pi_a^2 \ema ,
\qquad \bpi_a=\bm\bpi_a^1\cr\bpi_a^2 \ema .
\ee

In terms of these, 
\be
z_a=\pi_a^1/\pi_a^2,\qquad k_a\cdot k_b=\langle \pi_a,\pi_b\rangle[\bpi_a,\bpi_b],
\ee
with
\be
\langle \pi_a,\pi_b\rangle=\pi_a^1\pi_b^2- \pi_a^2\pi_b^1,\quad [ \bpi_a,\bpi_b]=\bpi_a^1\bpi_b^2- \bpi_a^2\bpi_b^1,\quad
z_a-z_b={\langle\pi_a,\pi_b\rangle\over\pi^2_a\pi^2_b}.
\ee
Then
\be 
\sum_{b\ne a}{k_a\cdot k_b\over z_a-z_b}=\sum_{b}\pi^2_a\pi^2_b[\bpi_a,\bpi_b]=0
\ee
by momentum conservation,
\be
\sigma_\mu \sum_{b}k_b^\mu= \sum_{b}\bpi_b\bpi_b^T=0.
 \ee 

Another solution to the scattering equations in $D=4$, not equivalent under M\"obius transformations to that of \cite{FR} (except for $N=4$), is given by $z_a=\bpi^1_a/\bpi^2_a$.
These are two rational explicit solutions which exist for all $N$. As noted in \cite{CHY3}, these solutions are can be associated with the MHV and anti-MHV amplitudes, and the others
can be associated other combinations of  helicities. [The scattering equations in four-dimensional space have also been discussed recently by Weinzierl \cite{SW}.] Although CHY introduced the scattering equations in order to describe tree amplitudes for Yang-Mills and 
gravity \cite{CHY1, CHY3}, perhaps more basically, they also describe scalar particles \cite{CHY2, DG1}. It seems that the other solutions can be associated with the division of the particles 
into $n$, say, positive helicities and $N-m$ negative helicities. However, perhaps strangely, the solutions only depend on how many positive and negative helicities there are, {\it i.e.} they depend on $n$, not on which 
particles are assigned positive helicity and which negative. 

To see this, as in \cite{DG2} (see also \cite{SV}), we divide the $N$ indices $a\in A$ into $m$ `positive' indices $i\in \P$ and $n$ `negative' indices $r\in \N$. Then the link variables $c_{ir}$, introduced by 
 Arkani-Hamed, Cachazo, Cheung and Kaplan \cite{ACCK1,ACCK2}, satisfy
\be
\pi_i=\sum_{s\in\N}c_{is}\pi_s,\qquad \bpi_r=-\sum_{j\in\P}\bpi_j c_{jr},\qquad i\in\P,\; r\in\N, \label{lv1}
 \ee
 and these are related to the parameters of twistor string theory \cite{W} -- \cite{BW} by 
\be
 c_{ir}=
 {\lambda_i\over\lambda_r (z_i-z_r)}, \qquad i\in\P,\; r\in\N.   \label{lv2}
 \ee
 Then, for $i\in\P$,
\begin{align}
\sum_a {k_i\cdot k_a\over z_i-z_a}&=\sum_r {\langle\pi_i,\pi_r\rangle [\bpi_i,\bpi_r]\over z_i-z_r}+\sum_j {\langle\pi_i,\pi_j\rangle [\bpi_i,\bpi_j]\over z_i-z_j}\\
\sum_r {\langle\pi_i,\pi_r\rangle [\bpi_i,\bpi_r]\over z_i-z_r}&=-\sum_{rsj} {c_{is}\langle\pi_s,\pi_r\rangle [\bpi_i,\bpi_j]c_{jr}\over z_i-z_r}
=\sum_{rsj} {\lambda_i\lambda_j\langle\pi_r,\pi_s\rangle [\bpi_i,\bpi_j]\over\lambda_r\lambda_s (z_i-z_r)(z_i-z_s)(z_j-z_r)}\cr
&=\half\sum_{rsj} {\lambda_i\lambda_j\langle\pi_r,\pi_s\rangle [\bpi_i,\bpi_j](z_r-z_s)\over\lambda_r\lambda_s (z_i-z_r)(z_i-z_s)(z_j-z_r)(z_j-z_s)}\cr
\sum_j {\langle\pi_i,\pi_j\rangle [\bpi_i,\bpi_j]\over z_i-z_j}&=\sum_{rsj}  c_{ir}c_{js}{\langle\pi_r,\pi_s\rangle [\bpi_i,\bpi_j]\over z_i-z_j}
=\sum_{rsj}  {\lambda_i\lambda_j\langle\pi_r,\pi_s\rangle [\bpi_i,\bpi_j]\over \lambda_r\lambda_s(z_i-z_j)(z_i-z_r)(z_j-z_s)}\cr
&=-\half\sum_{rsj}  {\lambda_i\lambda_j\langle\pi_r,\pi_s\rangle [\bpi_i,\bpi_j](z_r-z_s)\over \lambda_r\lambda_s(z_i-z_r)(z_j-z_s)(z_i-z_s)(z_j-z_r)},
\end{align}
where it is understood that $r,s$ are summed over $\N$ and $j$ over $\P$, with $j\ne i$. Thus the scattering equations (\ref{SE}) hold for $a=i\in\P$ and a similar argument shows that it holds if $a\in\N$.

 Spradlin and Volovich \cite{SV} conjectured that the number of solutions of the equations (\ref{lv1}) and  (\ref{lv2}) is given by the Eulerian number $\left\langle N-3\atop m-2\right\rangle, 2\leq m\leq N-2$, and CHY have given a demonstration of this \cite{CHY3}. These numbers sum to $(N-3)!$, the number of solutions of (\ref{SE}). This implies that the solutions depend on an assignment of a number of 
 positive and negative helicities, but must be independent of which particular momenta are associated with positive helicities and which negative, otherwise there would be too many solutions. To demonstrate this independence, consider moving $i$ into $\N$ and $r$ into $\P$,
\begin{align}
\pi_r={1\over c_{ir}}\pi_i-\sum_{s\ne r}{c_{is}\over c_{ir}} \pi_s&= {\lambda_r\over\lambda_i}(z_i-z_r)\pi_i-\sum_{s\ne r}{\lambda_r(z_i-z_r)\over \lambda_s(z_i-z_s)} \pi_s\cr
&= {\mu_r\over\mu_i(z_r-z_i)}\pi_i+\sum_{s\ne r}{\mu_r\over \mu_s(z_r-z_s)} \pi_s,
\end{align}
\begin{align}
\pi_j={c_{jr}\over c_{ir}}\pi_i+\sum_{s\ne r}\left[c_{js}-{c_{jr}c_{is}\over c_{ir}} \right]\pi_s
&={\lambda_j(z_i-z_r)\over \lambda_i(z_j-z_r)}\pi_i+\sum_{s\ne r} {\lambda_j\over\lambda_s}{(z_r-z_s)(z_j-z_i)\over (z_j-z_s)(z_i-z_s)(z_j-z_r)}\pi_s\cr
&={\mu_j\over \mu_i(z_j-z_i)}\pi_i+\sum_{s\ne r} {\mu_j\over\mu_s(z_j-z_s)}\pi_s
\end{align}
where
\be \lambda_r=
 {\mu_r\over z_r-z_i},\quad \lambda_i=\mu_i(z_i-z_r),\quad \lambda_s=
 \mu_s{z_s-z_r\over z_s-z_i},\quad \lambda_j=
 \mu_j{z_j-z_r\over z_j-z_i}\ee 
 So, defining $\tilde\P=\{r;j\in \P, j\ne i\}$ and $\tilde\N=\{i;s\in \N, s\ne r\}$, we have 
\be  \pi_\ell = \sum_{t\in \tilde\N}\tilde c_{\ell t}\pi_t,\quad \ell\in\tilde\P,\qquad \tilde c_{\ell t}={\mu_\ell\over \mu_t(z_\ell-z_t)};\ee 
and, similarly, we can show
\be  \bpi_t = \sum_{\ell\in \tilde\P}\bpi_\ell\tilde c_{\ell t},\quad t\in\tilde\N.\ee 
Thus, remarkably a solution $(z_a)$ to the scattering equations obtained from a division of the $N$ indices $a\in A$ into $n$ `positive' indices $i\in \P$ and $N-n$ `negative' indices $r\in \N$, does not depend on the choice of $\P, \N\subset A$, but only on $n$ [as explicitly seen above in the $n=2$ and $n=N-2$ `MHV'  and `anti-MHV' cases].

\section{Comments}
\label{Comments}

A general understanding of the polynomial form of the scattering equations was provided by the discussion of M\"obius invariance in section \ref{Moebius}. If the M\"obius algebra acts as in (\ref{L-1})--(\ref{L0}), the polynomial scattering equations are the only sets of homogeneous polynomials, related by M\"obius transformations, determining a finite set of points. 

In general, for a given highest weight polynomial $\varphi$, satisfying $L_{-1}\varphi =0, L_0\varphi=(\half N-n)\varphi$, the equations, $\varphi_m=0, n\leq m\leq N-n$, defined as in (\ref{defphim}), determine an algebraic variety in the complex projective space $\Cop\Pop^{N-1}$ on which the M\"obius group acts. [We shall assume $\varphi_S\ne 0, S\subset A, n\leq |S|\leq N-n$.] For $n=0$, where, up to a constant, $\varphi=1$, this is empty. For $1\leq n\leq \half N$, we can use the M\"obius invariance to send $z_1\rightarrow \infty, z_N\rightarrow 0$, as in (\ref{defh}) and (\ref{defhM}),
\be
h_m^\varphi=\lim_{z_1\rightarrow\infty}{1\over z_1} \varphi_{m+1}=\sum_{S\subset A'\atop |S|=m}\varphi_{S_1}z_S,\qquad
n-1\leq m\leq N-n-1,\label{defhphi}
\ee
where, again, $S_1= S\cup\{1\}$ and  $A'=\{a\in A: a\ne 1, N\}$. The homogeneous equations $h^\varphi_m=0, n-1\leq m\leq N-n-1,$ define a complex projective variety, $\V_\varphi$, in $\Cop\Pop^{N-3}$, which we expect typically to have dimension $2n-4$. For $n=1$, $h^\varphi_0$ is constant and there are no solutions, $\V_\varphi=\varnothing$. As we have discussed, $n=2$ provides precisely the scattering equations, for all possible light-like momenta satisfying momentum conservation, with typically $\V_\varphi$ containing $(N-3)!$ points. It would be interesting to determine conditions on $\varphi$ so that, for $n=2$, $\V_\varphi$ contains precisely this number of distinct points, and, for $n>2$, $\dim\V_\varphi=2n-4$. It would also be interesting to know whether the varieties $\V_\varphi$, for $n>2$, have any physical significance. 

We may realize differently the M\"obius group by changing the action of inversion from (\ref{inv}) to
\be
\phi(z_1,\dots, z_N)\mapsto (z_A)^K \phi(1/z_1,\ldots,1/z_N),\label{invK}
\ee
where $K$ is a positive integer, and then, in place of (\ref{L-1})--(\ref{L0}), which correspond to $K=1$, we have
\be
L_n=\sum_{a\in A}\left[-z_a^{n+1}{\partial\over \partial z_a}+K{n+1\over 2}z_a^n\right]
\ee
defining the action of translation, scaling and the special conformal transformation for $n=-1,0,1$, respectively.
In this case, for $N=1$, a single variable, the finite-dimensional invariant subspace has `M\"obius spin'  ${1\over 2}K$. 
For general $N$ it will be the tensor product of $N$ copies of this space and so will have dimension $(K+1)^N$; it consists of 
all polynomials involving no power of $z_a$ higher than $z_a^K$ for each $ a\in A$. Clearly, we may consider M\"obius invariant systems of 
 polynomial equations in this more general context but we have not yet analyzed these.

The CHY expressions for the tree amplitudes in scalar, gauge and gravity theories involve, in effect, summing a rational function (their integrand) over
the finite projective variety $\V_\varphi$, where $\varphi$ is the degree 2 highest weight polynomial whose coefficients are the scalar products of the external (light-like) momenta. Although the points of $\V_\varphi$ are, in general, irrational, the result is a rational function of the coefficients of $\varphi$ and the rational integrand. It would be desirable to have formula for this result, and an understanding of it, in terms of the algebro-geometric properties of $\V_\varphi$, perhaps along the lines of elimination theory \cite{GKZ}. The understanding of scattering equations  in terms of (\ref{p2}) and (\ref{defF}) may provide insight into possible extensions of the scattering  equations from the Riemann sphere to the torus. 

\section*{Acknowledgements}
We are grateful to Freddy Cachazo, David Fairlie and Song He for discussions.
LD thanks the Institute for Advanced Study at Princeton for its hospitality.
LD was partially supported by the U.S. Department of Energy, Grant No. 
DE-FG02-06ER-4141801, Task A, and PG was partially supported by NSF
grant No. PHY-1314311.
\vskip20pt
\singlespacing


\providecommand{\bysame}{\leavevmode\hbox to3em{\hrulefill}\thinspace}
\providecommand{\MR}{\relax\ifhmode\unskip\space\fi MR }
\providecommand{\MRhref}[2]
{
}
\providecommand{\href}[2]{#2}


\begin{thebibliography}{99}

\bibitem{CHY0}
F.~Cachazo, S.~He and E.Y.~Yuan,
{\it Scattering Equations and KLT Orthogonality,}
[\href{http://xxx.lanl.gov/abs/1306.6575}{{\tt arXiv:1306.6575}}].

\bibitem{CHY1}
F.~Cachazo, S.~He and E.Y.~Yuan,
{\it Scattering of Massless Particles in Arbitrary Dimensions,}
[\href{http://xxx.lanl.gov/abs/1307.2199}{{\tt arXiv:1307.2199}}].

\bibitem{CHY2}
F.~Cachazo, S.~He and E.Y.~Yuan,
{\it Scattering of Massless Particles: Scalars, Gluons and Gravitons,}
[\href{http://xxx.lanl.gov/abs/1309.0885}{{\tt arXiv:1309.0885}}].

\bibitem{DG1}
L.~Dolan and P.~Goddard,
{\it Proof of the Formula of Cachazo, He and Yuan
for Yang-Mills Tree Amplitudes in Arbitrary Dimension,}
[\href{http://xxx.lanl.gov/abs/1311.5200}{{\tt arXiv:1311.5200}}].

\bibitem{FR}
D.B.~Fairlie and D.E.~Roberts,
{\it Dual Models without Tachyons - a New Approach,} (unpublished Durham preprint PRINT-72-2440, 1972);
D.E.~Roberts, {\it Mathematical Structure of Dual Amplitudes,} (Durham PhD thesis, 1972) p.73 f. 
[\href{http://etheses.dur.ac.uk/8662/1/8662_5593.PDF}{{available at Durham E-Theses online}}]; 
D.B.~Fairlie, {\it A Coding of Real Null Four-Momenta into
World-Sheet Co-ordinates,} Adv. Math. Phys. {\bf 2009} (2009) 284689,
[\href{http://xxx.lanl.gov/abs/0805.2263}{{\tt arXiv:0805:2263}}].

\bibitem{GM} D.~J.~Gross and P.~F.~Mende, 
{\it String Theory Beyond the Planck Scale,}
Nucl.\ Phys.\ B {\bf 303}, 407 (1988).

\bibitem{Shaf}
I.R.~Shafarevich,
{\it Basic Algebraic Geometry,}
(Springer-Verlag, Berlin, 1974) p.198.

\bibitem{K}
C.~Kalousios,
{\it Massless Scattering at Special Kinematics as Jacobi
Polynomials,}
[\href{http://xxx.lanl.gov/abs/1312.7743}{{\tt arXiv:1312.7743}}].

\bibitem{GS}
J.J.~Sylvester, 
{\it On Derivation of Coexistence, Part II, being the Theory of 
Simultaneous Homogeneous Equations}, 
Philosophical Magazine {\bf 15} (1839) 428;
A.~Cayley,
{\it On the Theory of Elimination}, 
Cambridge and Dublin Mathematical Journal {\bf 3} (1848) 116;
G.~Salmon,
{\it Lessons Introductory to the Modern Higher Algebra,}
(Dublin, 1885) p.66.


\bibitem{CHY3}
F.~Cachazo, S.~He and E.Y.~Yuan,
{\it Scattering in Three Dimensions from Rational Maps,}
[\href{http://xxx.lanl.gov/abs/1306.2962}{{\tt arXiv:1306.2962}}].

\bibitem{SW} 
S.~Weinzierl, 
{\it On the Solutions of the Scattering Equations, }
[\href{http://xxx.lanl.gov/abs/1402.2516}{{\tt arXiv:1402.2516}}].

\bibitem{DG2}
L.~Dolan and P.~Goddard, 
{\it Gluon Tree Amplitudes in Open Twistor String Theory,}
JHEP {\bf 12} (2009) 032,
 [\href{http://xxx.lanl.gov/abs/0909.0499}{{\tt arXiv:0909.0499}}].



\bibitem{SV}
M.~Spradlin and A.~Volovich, 
{\it From Twistor String Theory To Recursion Relations,} 
Phys.Rev. {\bf D80} (2009) 085022, 
[\href{http://xxx.lanl.gov/abs/0909.0229}{{\tt arXiv:0909.0229}}].

\bibitem{ACCK1}
N.~Arkani-Hamed, F.~Cachazo, C.~Cheung and J.~Kaplan, 
{\it The S-Matrix in Twistor Space,}
[\href{http://xxx.lanl.gov/abs/0903.2110}{{\tt arXiv:0903.2110}}].

\bibitem{ACCK2}
N.~Arkani-Hamed, F.~Cachazo, C.~Cheung and J.~Kaplan, 
{\it  A Duality for the S Matrix,}
[\href{http://xxx.lanl.gov/abs/0907.5418}{{\tt arXiv:0907.5418}}].

\bibitem{W} 
E.~Witten, 
{\it Perturbative Gauge Theory as a String Theory in Twistor Space,} 
Commun. Math. Phys. {\bf 252} (2004)  189, 
[\href{http://xxx.lanl.gov/abs/hep-th/0312171}{{\tt arXiv:hep-th/0312171}}].

\bibitem{B} 
N.~Berkovits, 
{\it An Alternative String Theory in Twistor Space for N = 4 Super-Yang-Mills,}
Phys. Rev. Lett. {\bf 93} (2004) 011601,  
[\href{http://xxx.lanl.gov/abs/hep-th/0402045}{{\tt arXiv:hep-th/0402045}}].

\bibitem{BW} 
N.~Berkovits and E.~Witten, 
{\it Conformal Supergravity in Twistor-String Theory, }
JHEP {\bf 0408} (2004) 009,
[\href{http://xxx.lanl.gov/abs/hep-th/0406051}{{\tt arXiv:hep-th/0406051}}].

\bibitem{GKZ}
I.M.~Gel'fand, M.M.~Kapranov and A.V.~Zelevinsky,
{\it Discriminants, Resultants and Multidimensional Determinants,}
(Birkh\"auser, Boston, 1994).

\end{thebibliography}
\end{document}